# Efficient Switch Architectures for Pre-configured Backup Protection with Sharing in Elastic Optical Networks

Suthaharan Satkunarajah, Krishanthmohan Ratnam, and Roshan G. Ragel
Department of Computer Engineering
University of Peradeniya
Peradeniya, Sri Lanka.

*Abstract*— **In this paper, we address the problem of providing survivability in elastic optical networks (EONs). EONs use fine granular frequency slots or flexible grids, when compared to the conventional fixed grid networks and therefore utilize the frequency spectrum efficiently. For providing survivability in EONs, we consider a recently proposed survivability method for conventional fixed grid networks, known as pre-configured backup protection with sharing (PBPS), because of its benefits over the traditional survivability approaches such as dedicated and shared protection. In PBPS, backup paths can be pre-configured and at the same time they can share resources. Therefore, both short recovery time and efficient resource usage can be achieved. We find that the existing switch architectures do not support both PBPS and EONs. Specifically, we identify and illustrate that, if a switch architecture is not carefully designed, several key problems/issues might arise in certain scenarios. Such problems include unnecessary resource consumption, inability of using existing free resources, and incapability of sharing backup paths. These problems appear when PBPS is adopted in EONs and they do not arise in fixed grid networks. In this paper, we propose new switch architectures which support both PBPS and EONs. Particularly, we illustrate that, our switch architectures avoid the specific problems/issues mentioned above. Therefore, our switch architectures support using resources more efficiently and reducing blocking of requests.**

## I. Introduction

Optical networking with wavelength division multiplexing (WDM) has been considered to be a promising solution for handling the explosive growth of the Internet traffic [1]. WDM divides the vast transmission bandwidth available on a fiber into several non-overlapping wavelength channels and enables data transmission over the channels simultaneously. Typically, as specified by the International Telecommunication Union (ITU), 50 GHz or 100 GHz fixed grid spectrum spacing has been used for these channels in WDM. Recently, elastic optical (or flexgrid) networks (EONs) have received much attention for using the frequency spectrum more efficiently [2]. In EONs, fine granular frequency slots (12.5 GHz) or flexible grids are used for provisioning lightpaths instead of the conventional fixed grid spacing [3, 4]. The flexgrid technology allows assigning the spectrum according to bandwidth requirements and it enables expansion and contraction of lightpaths according to the traffic volume [5]. Therefore, the frequency spectrum is used more efficiently. To support EONs, two optical switch architectures have been proposed in the literature [6, 7]: gridless multi-granular and broadcast and select architectures which are shown in Fig. 1(a) and Fig. 1(b) respectively. In these architectures, components such as gridless/flexible/ bandwidth-variable wavelength selective switches (gridless/flex/BV WSSs) and bandwidth variable transceivers (BVTs) or gridless add/drop ports are primarily used to support EONs. In the gridless multi-granular architecture (shown in Fig. 1(a)), the gridless WSS is configured to select (switch) particular frequency slots. These frequency slots are then directed via the optical switch to the necessary output link. (In this architecture, a limited number of splitters are used for multicasting). Gridless add or drop ports are used to transmit or receive optical signals based on flexgrid. In the broadcast and select architecture, optical signals are broadcast (by splitters) to BV-WSSs attached to output links. Necessary frequency slots are then selected (switched) to the required output link by BV-WSSs.

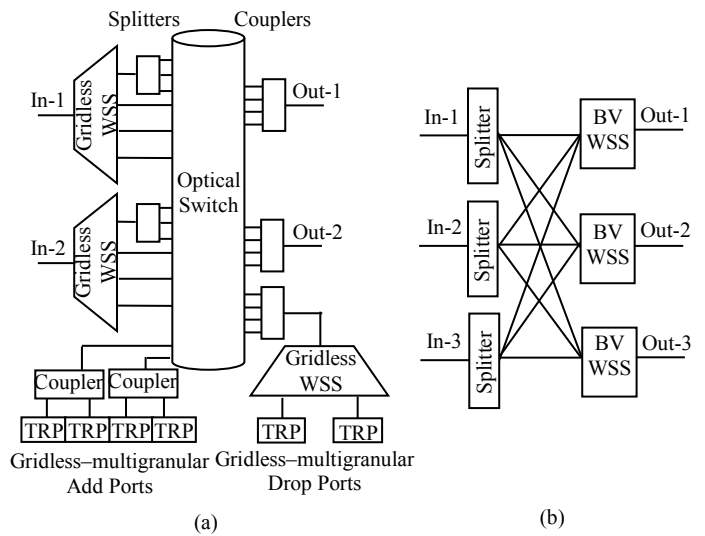

Fig. 1. Switch architectures for EONs (a) using gridless multi-granularity (b) using broadcast and select.

Providing survivability or protection with backup paths is an important issue in optical networks as a component failure such as a fiber-cut may disrupt large amount of multiplexed traffic. Traditionally, two survivability methods, dedicated and shared protection have been considered which are illustrated in Fig. 2(a) and Fig. 2(b) with a network topology of eight nodes and ten links. Solid arrows denote primary (P) or backup (B) paths which are pre-configured and dotted arrows denote backup (B) paths which are not pre-configured. In the dedicated protection shown in Fig. 2(a), backup paths B1 and B2 are configured at the time of establishing a connection (B1 and B2 are pre-configured). In case of a component failure, say, failure on P1, no further configuration is needed as B1 is pre-configured. Therefore, traffic is rerouted through B1 with a short recovery time. However, backup resources are not shared among backup paths and, therefore resource consumption in this approach is high [8] (six backup links used). In the shared protection shown in Fig. 2(b), backup paths B1 and B2 are not pre-configured. In this method, backup resources can be shared assuming the single link or component failure model, that is B1 and B2 share the wavelength on C-D, thus resources are used more efficiently (five backup links needed). However, it requires switch configuration to set up a backup path when a component failure occurs and it requires more recovery time: For instance, on a failure on P1, B1 must be configured or established before rerouting. Traditional switches based on techniques such as micro electromechanical systems (MEMS) do not support backup sharing while setting up pre-configured backup paths. Traditional survivability approaches have been investigated in EONs in [8-10].

Recently, a protection method, pre-configured backup protection with sharing (referred to as PBPS) has been proposed in [11] which is shown in Fig. 2(c). This method uses a new switch architecture [11, 12] shown in Fig. 3 which supports backup sharing while setting up pre-configured backup paths. In Fig. 2(c), B1 and B2 are pre-configured while they share the wavelength on C-D. The architecture used for PBPS [11, 12] uses variable optical splitters (VOSs), which can be configured to vary the power splitting ratio on output ports. Therefore, VOSs can be used to direct the input power towards one output port only (like the traditional switches). Further, they can also be used to split the power on a desired sub-set of output ports on need basis (details given in Section III). Combiners are also used in the architecture. Using VOSs and combiners, backup sharing can be done while setting up pre-configured backup paths. That is, the switch at node C can be configured to connect two (or more) input ports or links A-C and E-C to the same output port C-D at the same time. With this pre-configuration, the traffic can be switched from one of the input ports, either from B1 or B2 to the same output port C-D which requires no further configuration. Further, the switch at node D can be pre-configured to split power from C-D on two output ports D-B and D-F. With this pre-configuration, the traffic, either from B1 or B2 can be switched on one of the output ports with the split power, either to D-B and D-F, which requires no further configuration. Therefore, in PBPS, the benefits gained in both dedicated and shared protection methods such as short recovery time (from pre-configured backup paths) and efficient resource usage (from backup sharing) are achieved. In [11], the PBPS has been investigated for fixed grid networks only. The switch architecture used for PBPS [11, 12] (Fig. 4) does not support EONs.

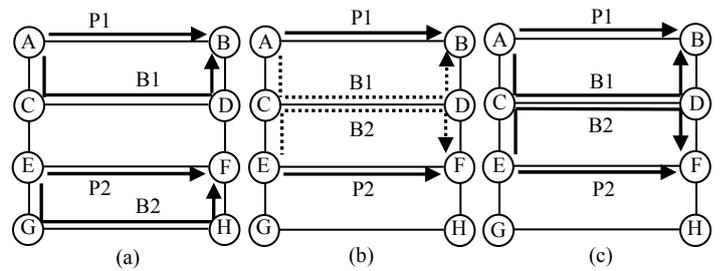

Fig. 2. Protection methods (a) Dedicated (b) Shared (c) Pre-configured backup protection with sharing (PBPS)

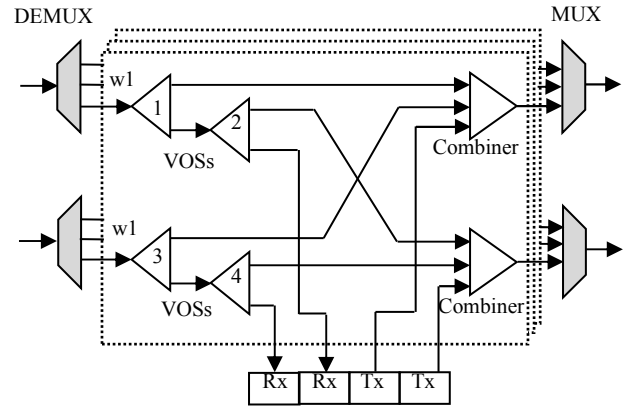

Fig. 3. Switch architecture with VOSs for fixed grid networks

In this paper, we focus on EONs and the PBPS survivability method, because of the benefits mentioned above. To the best of our knowledge, this is the first paper that is investigating the PBPS approach in EONs. A fundamental issue in this investigation is that the optical switches should support both PBPS and EONs. The existing switch architectures proposed for EONs [6, 7] do not directly support the PBPS approach (details given in Section II). Therefore, our objective is to propose switch architectures, which supports both PBPS and EONs. Specifically, we identify the following problems which might arise in some scenarios if a switch architecture is not carefully designed for PBPS and EONs: (1) some frequency slots or bandwidth resources might be consumed unnecessarily, (2) some existing free frequency slots cannot be used by future requests, (3) backup sharing cannot be done, even though primary paths are link disjoint and enough free frequency slots are available (details are given in Section II). These problems pose challenges in designing a switch architecture for supporting both PBPS and EONs. These problems are seen in (or related to) scenarios in which pre-configured backup paths are shared with partially overlapping frequency slots. They lead to high resource consumption or more blocking of requests. These problems appear when PBPS is adopted in EONs and they do not arise in fixed grid networks. Further, they do not arise when the traditional survivability approaches such as dedicated and shared protection are adopted in fixed or flexgrid networks.

In this paper, we propose new optical switch architectures which support both PBPS and EONs. More importantly, our proposed switch architectures avoid the specific problems mentioned above in circuit level traffic. However, we note that one of our proposed switch architectures will also support bursty level traffic. This is an additional feature supported by the switch other than supporting PBPS and EONs.

## II. SUPPORTING PBPS IN EONs AND POTENTIAL ISSUES

In this section we illustrate why existing switches do not directly support the PBPS approach. Specifically, with a slightly modified architecture from one of the existing architectures, we illustrate the potential problems or issues that arise in some scenarios when the PBPS approach is adopted in EONs. We consider the PBPS illustration shown in Fig. 2(c) and particularly we consider the required switch configurations at node C and D.

The gridless multi-granular architecture [6] (Fig. 1(a)) proposed for EONs consists of an optical switch, gridless WSSs, transponders, couplers and splitters. We assume that traditional optical switches such as MEMS are used in this architecture. With the couplers and the optical switch in this architecture, similar switch configuration at node C as explained in Section I (Fig. 2(c)) can be done to connect or switch certain frequency slots from A-C and E-C to C-D. However, to have similar switch configuration at node D as explained in Section I, that is splitting (Fig. 2(c)), splitters are necessary. In the existing architecture shown in Fig. 1(a), only a limited number of splitters are used for multicasting purposes. Therefore, PBPS is not supported if enough splitters are not available. Even if a slightly modified switch architecture with more splitters as shown in Fig. 5 is considered, several problems or issues appear in certain scenarios though it supports the basic configurations required for PBPS. We illustrate these problems below.

To illustrate the problems, we consider a network topology with six nodes and seven links as shown in Fig. 4(a). Primary paths P1 and P2 and their associated backup paths B1 and B2 are shown with their frequency slots 1-10 and 1-3 respectively (Note that it is possible that a primary and its backup can consume different frequency slots). For illustration, we do not consider guard bands. The required (or preferred) frequency slots on links for these primary and backup paths are shown in square-brackets. In this network, node C is configured for PBPS as explained above, hence, link C-D is shared with partially overlapping frequency slots as B1 and B2 consume 1-10 and 1-3 respectively (Fig. 4(b)(i)). The other scenarios of partially overlapping frequency slots with B2 consuming different frequency slots are also possible, such as B1 and B2 consuming frequency slots 1-10 and 4-6 (Fig. 4(b)(ii)) and 1-10 and 9-11 (Fig. 4(b)(iii)). We consider the switch configuration at node D with the slightly modified switch architecture (from Fig. 1(a)) shown in Fig. 5. With this configuration, we illustrate the following problems which arise in the scenario shown in Fig. 4(a). Note that, these problems arise in other scenarios shown in Fig. 4(b) as well. In addition to this, similar problems appear when three or more backup paths share a link with partially overlapping frequency slots.

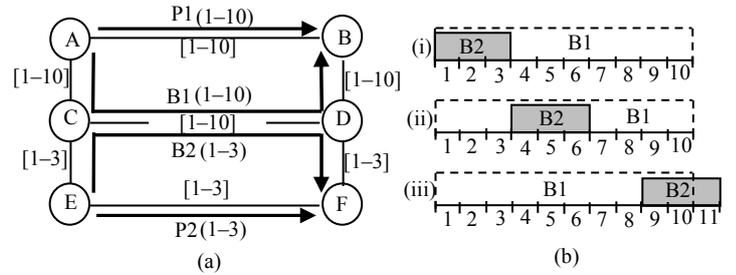

Fig. 4: (a) A scenario of PBPS with partially overlapping frequency slots (b) Possible scenarios of partially overlapping frequency slots on link C–D

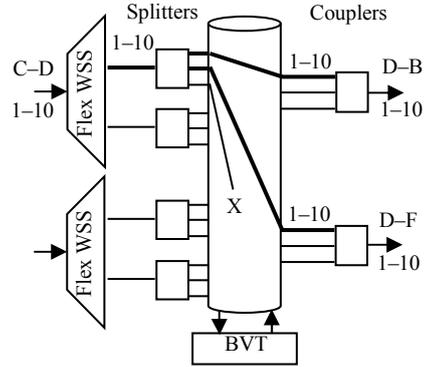

Fig. 5. Switch configuration at node D with a slightly modified switch architecture from Fig. 1(a)

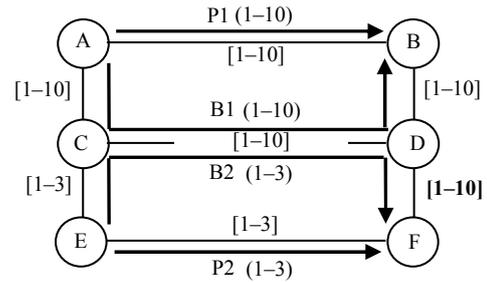

Fig. 6. Unnecessary resource usage on link D–F

### A. Unnecessary consumption of resources

The modified switch is configured for PBPS at node D as shown in Fig. 5 in which flex WSS on the input link C-D, selects (switches) the frequency slots 1-10 (consumed by B1-fully and B2-partially) and directs them to the optical switch via the splitter. The optical switch, directs these frequency slots to the couplers. Since, the couplers are not able to selectively direct the necessary frequency slots which are 1-10 and 1-3 to the output links, the frequency slots 1-10 are consumed by B2 on link D-F instead of 1-3 as shown in Fig. 6. Thus, frequency slots 4-10 are consumed additionally on D-F.

### B. Future requests cannot use some existing free resources

For illustration, we suppose that, the architecture shown in Fig. 5 is further modified with more components such that, it avoids unnecessary consumption of resources or frequency slots as explained above. That is, B2 now consumes 1-3 on D-F instead of 1-10. Details of such additional components are given in Section III. Though, frequency slots 4-10 are available on D-F, future requests cannot use these existing free resources. To illustrate this problem, we consider that the frequency slots 4-10 are consumed by a new request P3, which is set up from D to F (not shown). This request is added from BVT (Fig. 5) and it should be directed through the coupler on link D-F. Note that, the same coupler has already been used to direct B2 with 1-10 (Fig. 5). If there is a failure on P1 (Fig 4(a)), traffic will be rerouted through B1 with 1-10. Because of the splitter at node D, copy of traffic will also be directed on D-F via B2 (Fig. 5) through the coupler. This traffic will interfere with the new request P3's traffic at this coupler, because both their frequency slots (1-10 and 4-10) overlap. Therefore, the architecture does not support P3 in this scenario. This illustrates that, free resources or frequency slots cannot be used by new requests.

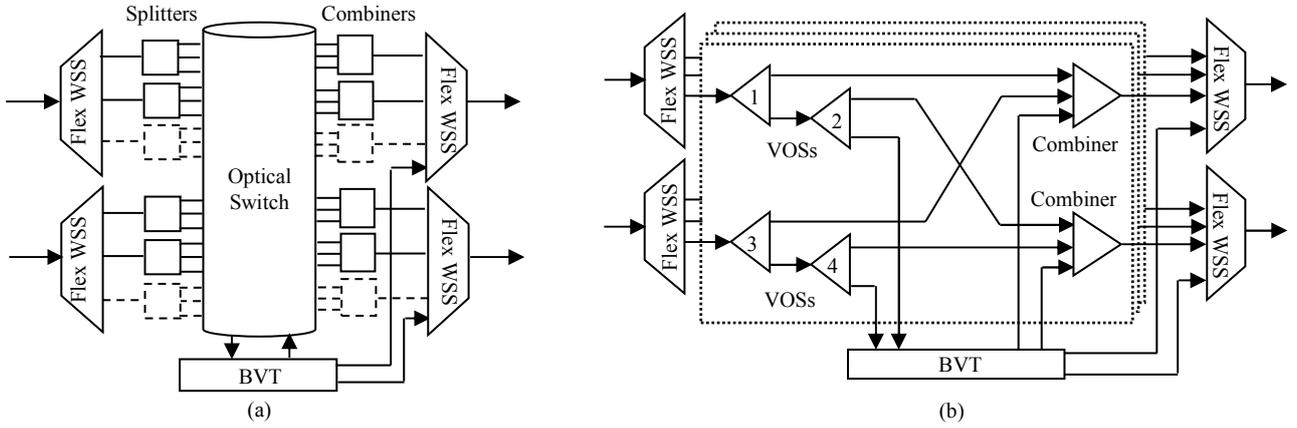

Fig. 7. Switch architectures for PBPS in EONs (a) using traditional optical switches (b) using VOSs

*C. Backup sharing in PBPS is not possible in certain cases*

We explain this problem by considering the order of request arrival for P1, P2, and P3 (considered above) differently. Suppose that, of the three requests, P3 (from D-F) arrives first and established as explained above. Then P1 arrives, and it is set up with its backup B1 as shown in Fig. 4(a). Note that as P2 has not arrived yet, optical switch (Fig. 5) only directs 1-10 from one of splitter's outputs to D-B and it simply blocks all other outputs from the splitter. The problem is that, when P2 arrives, its backup B2 cannot be set up according to PBPS as shown in Fig. 4(a). Because, if it is set up, the same interference problem as explained above appears. Because, to set up B2, optical switch should now direct the second output from the splitter towards D-F and the rerouted traffic from C-D (on failure) will interfere with P3. Therefore, B2 cannot be established with sharing in PBPS in this case.

The broadcast and select architecture [7] shown in Fig. 1(b) has splitters and BV-WSSs: because of the splitters, similar switch configuration at node D (Fig. 2(c)) as explained in Section I, that is splitting, can be done for PBPS in EONs. However, it appears that, switch configuration at node C (combining, as explained before) is not possible with this architecture. Because it appears that BV-WSS with such as liquid crystals on silicon (LCOS) or MEMS do not support the configuration to select the same frequency slots from two input ports and switching them on the same output ports at the same time. This is the problem in this architecture to support PBPS in EONs.

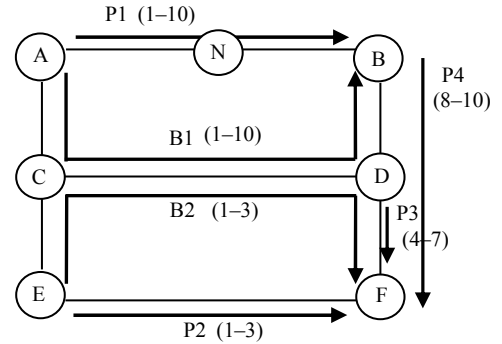

Fig. 8. A scenario of PBPS

### III. PROPOSED OPTICAL SWITCH ARCHITECTURES

We propose two optical switch architectures for supporting PBPS in EONs: (a) switch architecture for PBPS in EONs using traditional optical switches, and (b) switch architecture for PBPS in EONs using VOSs. Particularly, our proposed architectures avoid the problems or issues illustrated in Section II. Our proposed switch architectures with two input links and two output links are shown in Fig. 7. Our first switch architecture, shown in Fig 7(a), uses the traditional optical switch. Our second switch architecture, shown in Fig 7(b), does not use the traditional switch. Therefore, the advantage of using the first switch is that the existing switches could be added with additional components as shown in Fig. 7(a) to support PBPS in EONs. However, our second switch supports bursty level traffic in addition to circuit level traffic. This feature (bursts) is not supported by the first switch because of slow traditional switches. Further, unlike fixed splitters, VOSs used in our second architecture avoid unnecessary power wastage as power can be split only on a subset of ports as we needed. In addition to this, instead of using separate components such as splitters and traditional optical switches, VOSs are used in this architecture, which can perform both splitting and switching functions. These are the advantages of using our second architecture. Below we describe each architecture and illustrate how PBPS is supported in EONs. Particularly we illustrate how they avoid the problems mentioned in Section II.

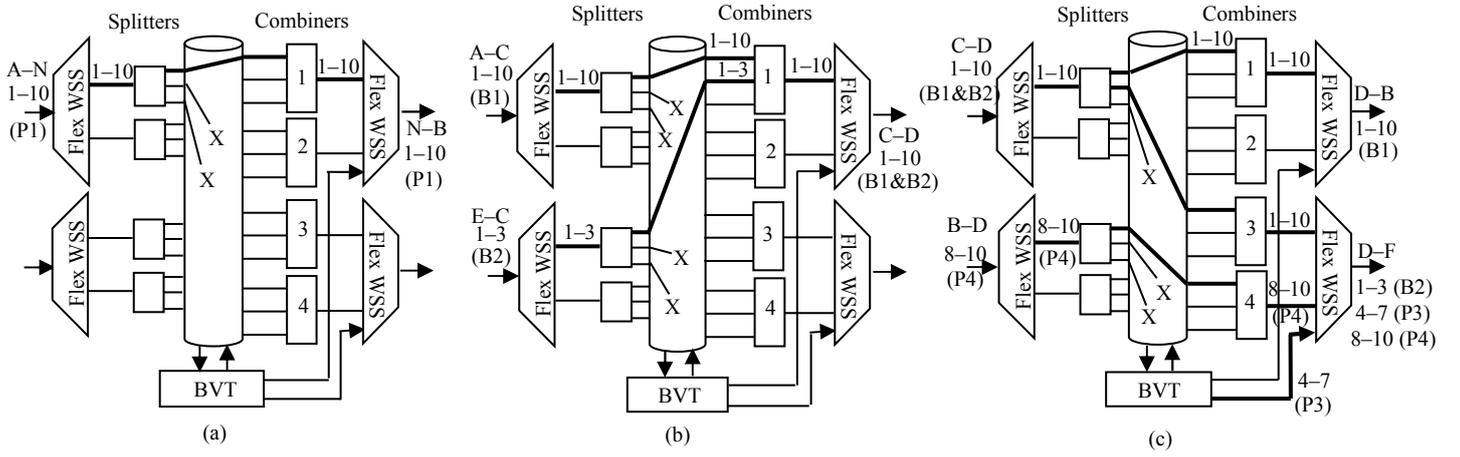

Fig. 9. Switch configuration with the architecture shown in Fig. 8(b) at (a) node N (b) node C and (c) node D

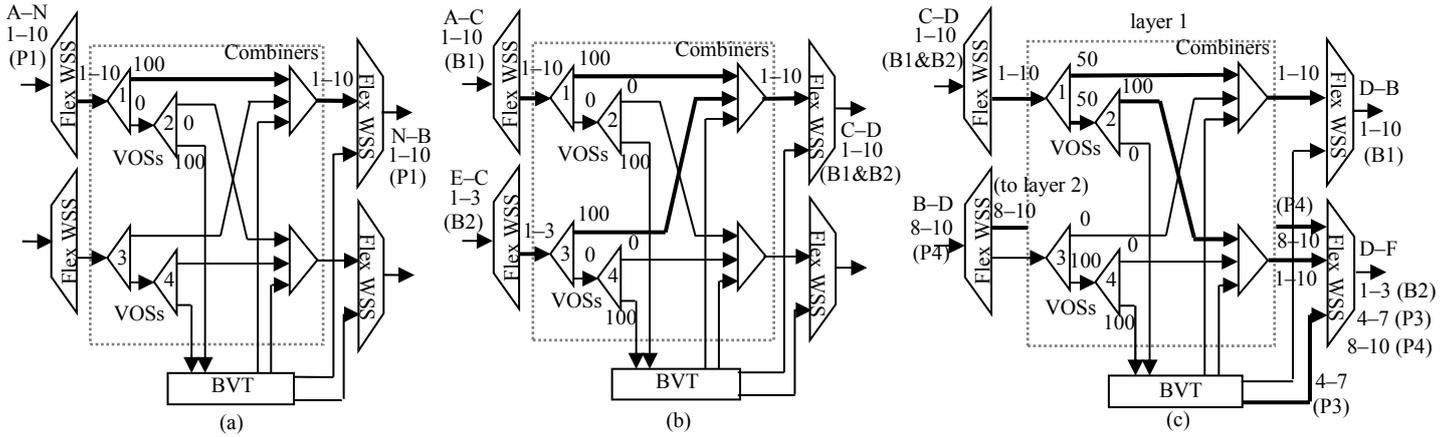

Fig. 10. Switch configuration with the architecture shown in Fig. 8(b) at (a) node N (b) node C and (c) node D

*A. Switch architecture for PBPS in EONs using traditional optical switches*

The switch shown in Fig. 7(a) is a modified architecture from the switch shown in Fig. 1(a) [6]. In this architecture, flex WSSs are used to perform functionalities such as demultiplexing/multiplexing and switching variable spectral bandwidths using integrated spatial optics in EONs. These devices are based on one of several technologies: MEMS or LCOS [13]. We use BVT to generate an optical signal using just enough spectral resources in EONs [14]. We include additional splitters to enable PBPS in EONs. Further, flex WSSs are included on the output links and additional combiners and new direct connections from BVT to flex WSSs are also included in this architecture. These new additions are necessary to avoid potential problems illustrated in Section II.

*1) Setting up primary and backup paths in PBPS:* A network topology with four primary paths P1, P2, P3 and P4 and two backup paths B1 (for P1) and B2 (for P2) are shown in Fig. 8 (we assume that the link B-D is bi-directional and all other links are unidirectional). With our switch architecture shown in Fig. 7(a), switch configurations at node N (to illustrate how a typical switching is done), at node C (to illustrate PBPS configurations) and at node D (to illustrate PBPS configurations and how potential problems are avoided) are shown in Fig. 9. Switch configurations at nodes A, B, E and F for add/drop operations are not shown due to space constraints. We use the 2x2 switch architecture shown in Fig. 7(a) for these illustrations.

At node N, P1 is switched from A-N to N-B (Fig. 9(a)). Flex WSS switches the frequency slots 1-10 to the splitter. Optical switch directs the selected frequency slots (1-10) from the splitter to the flex WSS on the output link via the combiner. Switch configurations at node C and node D for PBPS with B1 and B2 (combining and splitting) are shown in Fig. 9(b) and Fig. 9(c). In Fig. 9(b), frequency slots 1-10 and 1-3 for B1 and B2 are selected from splitters and they are directed to the combiner. Flex WSS on the output link then filters the frequency slots 1-10 for B1 and B2. Note that only one of B1 and B2 will be active with rerouted traffic (on failure) at a time under the single components failure model. At node D (Fig. 9(c)), the optical switch directs the frequency slots 1-10 from C-D (from splitter's output ports) to the combiners connected to D-B and D-F. Each flex WSS on the output links filters 1-10 (B1) and 1-3 (B2).

*2) Avoiding potential problems:* We consider the problem of unnecessary consumption of resources illustrated in Section II. To avoid this problem, flex WSSs are included on the output links of our switch as shown in Fig. 7(a). The flex WSS on the output link D-F filters the required frequency slots 1-3 (B2) as shown in Fig. 9(c). Therefore, B2 does not consume frequency slots 4-10 on D-F, hence, unnecessary resource or frequency slot consumption on D-F is avoided.

We now consider the second problem, that is free resources cannot be used by future requests (as explained in Section II) even though frequency slots are available on the output links (by including flex WSSs). This problem is avoided by (a) adding direct connections from BVTs to flex WSSs and (b) including more combiners in our switch architecture as shown in Fig. 7(a). We consider the frequency slots 4-10 which are available on D-F for future use as the flex WSS switches only relevant frequency slots as explained above. We illustrate how these frequency slots 4-10 can be used by considering two requests P3 and P4 as shown in Fig. 8. Note that these available frequency slots 4-10 on D-F can be used by two possible ways: by locally-added-traffic and switched-traffic. P3 is locally added at D which consumes 4-7 slots and P4 is switched at D (bypass traffic) which consumes 8-10 slots. P3 is set up from D to F. To add this request, the direct connection from BVT to flex WSS is used (shown in bold line in Fig. 9(c)). In addition, P4 is set up from B to F passing through node D. To switch P4 at D, the flex WSS on B-D directs frequency slots 8-10 to the optical switch via the splitter. The optical switch then directs these frequency slots to the combiner-4 (Fig. 9(c)). As P3 is set up through the direct connection from BVT to flex WSS and P4 is set up through the separate combiner (B2 uses combiner-3 whereas P4 uses combiner-4), there will not be any interference problem at the combiner as explained in Section II.

We consider the third problem, that is backup sharing in PBPS is not possible in certain cases as explained in section II. This problem is also avoided by (a) adding direct connections from BVTs to flex WSSs and (b) adding more combiners in our switch architecture (Fig. 7(a)). As explained above, B2 and P4 are using different combiners and P3 does not use combiners at D as it is set up with the direct connection to avoid interference. Therefore it is clear that, backup sharing of B1 and B2 is possible, even though the requests arrive in different order.

### B. Switch architecture for PBPS in EONs using VOSs

Our switch shown in Fig. 7(b) uses components such as VOSs and combiners as used in the architecture proposed in [11] (Fig. 3) and includes new components such as flex WSS and BVT. Further, new direct connections from BVTs to flex WSSs are included in addition to connections from BVT to combiners. Components such as VOSs and combiners are integrated in layers as shown in the figure. We avoid the potential problems by having these direct connections and multiple layers.

We use the 1x2 VOS component presented in [15-17]. The self-latching VOS is based on magneto-optical technology. The VOS is designed using a variable faraday rotator and a walk-off crystal. Switching is done using the power directing and power splitting functionalities of the VOS. In VOS, input optical power can be split on the two output ports with various ratios (states) such as (0% -100%), (50% -50%), and (100%-0%). The states (0%-100%) and (100%-0%) can be used to direct the full power on an output port. It takes 0.25ms time to switch between (0%-100%) and (50%-50%) states. Average insertion loss (IL) is 0.6dB (for (0%-100%) and (100%-0%) states) and 4dB (for (50%-50%) states), and polarization-dependent loss (PDL) is less than 0.1dB. The VOS energy consumption is very low (~120μJ) [16]. The high speed VOSs available in [18] can also be used in our architecture.

*1) Setting up primary and backup paths in PBPS:* With our switch architecture shown in Fig. 7(b), switch configurations (for Fig. 8) at node N (to illustrate how a typical switching is done), at node C (to illustrate PBPS configurations) and at node D (to illustrate PBPS configurations and how potential problems are avoided) are shown in Fig. 10.

Switch configuration for P1 at node N is shown in Fig. 10(a). Flex WSSs are configured as explained in our first architecture and VOS-1 is configured for (100%-0%) as shown to direct full power towards node B. Switch configurations at node C and node D for PBPS with B1 and B2 (combining and splitting) are shown in Fig. 10(b) and Fig. 10(c). In Fig 10(b), VOS-1 and VOS-3 are configured for (100%-0%) to direct full power towards link C-D for PBPS with B1 and B2. At node D (Fig. 10(c)), VOS-1 and VOS-2 are configured for (50% -50%) and (100%-0%) states. With these configurations, frequency slots 1-10 on link C-D can be switched towards links D-B and D-F and flex WSSs on the output links select 1-10 and 1-3 and switch these on D-B and D-F respectively.

*2) Avoiding potential problems:* We consider the problem of unnecessary consumption of resources as illustrated in Section II. With flex WSSs on output links (Fig. 7(b)), they can selectively switch necessary frequency slots as explained in our first architecture and avoid this problem (shown in Fig. 10(c)). We consider the second problem that is free resources cannot be used by future requests (as explained in Section II) even though frequency slots are available on the output links. This problem is avoided by (a) including additional direct connections from BVTs to flex WSSs and (b) having layers which contains more combiners (Fig. 7(b)). As shown in Fig. 10(c), P3 is set up through the direct connection from BVT to flex WSS (shown in bold line) and P4 is set up through the second layer (using separate combiner and VOSs). Hence, there is no interference problem as explained above. They use free resources (4-10) on D-F and thus they avoid this problem. The third problem that is backup sharing in PBPS is not possible in certain cases (as illustrated in Section II) is also avoided with (a) additional connections from BVTs to flex WSSs and (b) combiners in separate layers as explained in our first architecture.

As explained above in the two architectures, primary paths are added through direct connections from BVTs to flex WSSs to avoid interference. Backup paths are added through combiners (combiners in our first traditional optical switch based architecture, and combiners in layers in our second VOS based architecture) to increase the possibility of backup sharing. In this paper, we use the switches shown in Fig. 7 for supporting circuit level switching. In addition, our VOS based architecture (Fig. 7(b)) supports bursty level switching also, in which flex WSSs are to be pre-configured for certain range of frequency slots for bursts. Bursts arriving on certain frequencies can be switched using VOSs with the switching method proposed in [12]. We note that, the architecture based on traditional switches (Fig. 7(a)) and the existing switch architecture shown in Fig. 1(a) do not support bursty level traffic (burst switching), because of slow switches such as MEMS (Fig. 1(a)) and high response or switching time of flex WSS (300 ms [19]). In the architecture shown in Fig. 1(b), flex WSSs are the switching components.

## IV. CONCLUSIONS

In this paper, we addressed the problem of providing survivability using pre-configured backup protection with sharing (PBPS) in elastic optical networks (EONs). We considered circuit level switching in this paper. Our contributions are twofold. Firstly, we identified several potential problems which might arise in certain scenarios when PBPS is adopted in EONs. These problems cause high resource consumption or more blocking of requests. Secondly, we proposed two switch architectures that support both PBPS and EONs. Particularly, our architectures eliminate these potential challenges. Our first architecture uses traditional switching components. Therefore, the advantage is that the existing switches could be added with additional components according to our architecture to support PBPS and EONs. Our second architecture does not use the traditional switching components. However, it supports an additional feature, among other benefits, which is supporting bursty level switching. In our future work, we will quantitatively analyze the performance of our switch architectures for resource usage and blocking performance, using simulation experiments.